\documentclass[graphics,floatfix, footinbib,tightenlines,nobibnotes, aps, prb, twocolumn]{revtex4-1}
\usepackage{amsmath,amssymb,wasysym}
\usepackage{graphicx}
\usepackage{dcolumn}
\usepackage{bm}
\usepackage{braket}
\usepackage{subcaption}
\usepackage{verbatim}
\usepackage{float}
\usepackage{bbold}
\usepackage{color}
\usepackage{xcolor}
\usepackage{relsize}
\usepackage{amsthm}
\usepackage{enumerate}
\usepackage{soul,xcolor}
\usepackage[normalem]{ulem} 

\usepackage[T1]{fontenc}
\usepackage[colorlinks=true ,urlcolor=blue,urlbordercolor={0 1 1}]{hyperref}

\newcommand{\beq}{\begin{eqnarray}}
\newcommand{\eeq}{\end{eqnarray}}

\newcommand{\bsp}{\begin{split}}
\newcommand{\esp}{\end{split}}

\newcommand{\be}{\begin{equation}}
\newcommand{\ee}{\end{equation}}

\newcommand{\moire} {moir{\' e} }

\begin{document}

\setstcolor{red}

\title{Landau Level Degeneracy in Twisted Bilayer Graphene: Role of Symmetry Breaking}
\author{
Ya-Hui Zhang, Hoi Chun Po, T. Senthil}
\affiliation{Department of Physics, Massachusetts Institute of Technology, Cambridge, MA, USA
}

\date{\today}

\begin{abstract}

The degeneracy of Landau levels flanking charge neutrality in twisted bilayer graphene is known to change from eight-fold to four-fold when the twist angle is reduced to values near the magic angle of $\approx 1.05^\circ$. This degeneracy lifting has been reproduced in experiments by multiple groups, and is known to occur even in devices which do not harbor the correlated insulators and superconductors.
We propose $C_3$ symmetry breaking as an explanation of such robust degeneracy lifting, and support our proposal by numerical results on the Landau level spectrum in near-magic-angle twisted bilayer graphene.
Motivated by recent experiments, we further consider the effect of $C_2$ symmetry breaking on the Landau levels.
\end{abstract}

\pacs{Valid PACS appear here}
\maketitle

\section{Introduction}
The discovery\cite{cao2018correlated, cao2018unconventional} of correlated insulators and superconductivity in magic-angle twisted bilayer graphene (TBG) has sparked tremendous experimental and theoretical activity. The original results of Refs.  \onlinecite{cao2018correlated, cao2018unconventional} have been confirmed and extended in Refs. \onlinecite{yankowitz2019tuning,lu2019superconductors}. 
Ferromagnetism, accompanied by an anomalous (possibly quantized) Hall effect, has also been observed at 3/4 filling of the conduction band for some devices\cite{Aaron2019Emergent, lu2019superconductors}. 
In addition, gate tunable correlated insulators\cite{Wang2019Evidence} as well as signs of superconductivity\cite{Wang2019Signatures} have been demonstrated in the moire bands of ABC trilayer graphene aligned with a hexagonal boron nitride (h-BN) substrate. Very recently, twisted double bilayers of graphene have also been studied and are shown to host spin-polarized correlated insulators\cite{Shen2019Observation, Liu2019Spin,Cao2019Electric}  and superconductivity\cite{Shen2019Observation, Liu2019Spin}.

In spite of the experimental progress, theoretically there is very little understanding of the many-body physics of these systems. In this paper, we address one aspect of the phenomenology of the normal metallic state of near-magic-angle twisted bilayer graphene that has resisted explanation so far. Specifically, we will focus on understanding the Landau level degeneracy near charge neutrality.  As we discuss below, the same pattern of Landau fan is consistently observed across samples with varying twist angles and in a reasonably wide range of densities and temperatures. This robust experimental observation, however, has resisted a clear theoretical explanation so far.

More specifically, it is well known that the band structure of twisted bilayer graphene features two Dirac points at charge neutrality within each microscopic valley. This band touching is protected by a $C_2 \mathcal T$ symmetry\footnote{$C_2$ refers to 180 degree spatial rotation and $\mathcal T$ to time reversal}. We will refer to these as mini-Dirac points.  
With two mini-Dirac points per valley and per spin, we have a total of eight Dirac points, which is double of that in monolayer graphene.
Upon doping away from neutrality, Landau levels will form out of the mini-Dirac cones. The corresponding Landau fan sequence will then be expected to be double of that of monolayer graphene, namely, $\pm4, \pm12, \pm 20, \dots$. Indeed, precisely this degeneracy pattern is seen in experiments on devices with relatively large twist angle of $1.8^\circ$ \cite{Cao2016Superlattice}, which is far away from the magic angle $\approx 1.1^\circ$.

In the vicinity of the magic angle, however, there is a surprise. The Landau fan emerging from neutrality has the Hall conductance sequence $\pm 4, \pm 8, \pm 12, \dots$.
In other words, the eight-fold Landau-level degeneracy is reduced to four-fold.  
Experimentally, this degeneracy lifting is always seen in samples showcasing the correlated insulators and superconductivity. However, the converse is not true, i.e., the same sequence is observed even in some devices which do not show the other correlated phenomena \footnote{
For instance, see supplementary data in Ref.\  \onlinecite{yankowitz2019tuning} on the $1.27^\circ$ device (D2) at the ambient pressure of $0$ GPa.
}.
Furthermore, the Landau fans are found to terminate once the half-filling correlated insulators set in, and the degeneracy is further reduced to two-fold on the other side of the insulator. As superconductivity is seen on both sides of the insulators, understanding the Landau fan, which conveys information on the nature of the charge carriers and the possible patterns of symmetry breaking, may shed light on the nature of the correlated phenomena closer to the magic angle.

Here, we show that the experimentally observed Landau fan can be reproduced within a free-fermion model of twisted bilayer graphene, for various angles close to the magic angle.
Our main result is that the sequence is stabilized by a weak breaking of $C_3$ symmetry.
As is well-known, $C_3$ symmetry breaking splits the Van Hove singularity (VHs) in the density of states \cite{choi2019imaging, Bi2019Designing}.
For concreteness, let us focus on the conduction band.
In the energy range between the two split Van Hove singularities, the equal-energy contours in the momentum space (within each valley) take a qualitatively different shape compared to those allowed in the symmetric case: there is a single electron-like orbit which encloses both of the Dirac points. In contrast, without $C_3$ symmetry breaking, the equal energy contours either consist of  two disjoint pockets each enclosing  a  Dirac point, or a single hole-like pocket enclosing the band maximum at $\Gamma$.  
In the presence of a magnetic field, these new equal-energy contours give rise to four-fold degenerate Landau levels stemming out from charge neutrality. 
We find that even for a relatively small degree of the $C_3$ breaking, the degeneracy in the vicinity of the magic angle is four-fold at the magnetic field strengths at which the experiments are done ($\gtrapprox 1$ T) .

We trace the origin of this effect to the large susceptibility of near-magic-angle TBG to such a $C_3$ breaking perturbation, which leads to a significant modification to the electronic spectrum even when the ``bare'' $C_3$ breaking is weak.
Such symmetry breaking may arise due to strain effects (known to be generally present in TBG), or due to spontaneous symmetry breaking driven by interactions. While we leave open the physical origin of the $C_3$ breaking, we remark that the smallness of symmetry breaking required suggests that the presence of strain in the sample is a plausible explanation.

In some devices of twisted bilayer graphene, for example in Ref. \onlinecite{Aaron2019Emergent}, it is known that there is very likely also a breaking of $C_2$ symmetry. For the device in Ref. \onlinecite{Aaron2019Emergent}, this is due to the near alignment with a h-BN substrate. Such $C_2$ breaking gaps out the Dirac points and results in an insulator at charge neutrality. In the device studied in Ref. \onlinecite{lu2019superconductors}, it is observed that the system is insulating at neutrality. A possible explanation is that $C_2$ is broken in this system as well, although it is unclear if the symmetry breaking is again due to alignment with h-BN, or is interaction driven 
In light of these experimental results, we also study the expected Landau levle degeneracy in  devices with $C_2$ symmetry breaking, both with and without an additional $C_3$ breaking.

We remark that multiple theoretical attempts have already been made to address the neutrality Landau fan  in small-angle TBG \cite{Bistritzer2011PRB, deGail2011Topologically, Moon2012Energy, lian2018landau, Hejazi2019Landau}. While all of the emergent symmetries of TBG are preserved in these earlier works, some of them did identify specific choice of parameters for which the experimental sequence of $\pm 4, \pm 8, \pm 12, \dots$ can be observed. For instance, Ref.\ \onlinecite{Hejazi2019Landau} found numerically that the experimental sequence can be reproduced in a narrow range of parameters near the magic angle. However, the sequence is also found to depend very sensitively on the choice of model parameters \cite{Hejazi2019Landau}, and might not explain the experimental robustness (across groups and samples) of the Landau level sequence.

Alternatively, Ref.\ \onlinecite{deGail2011Topologically} proposed a VHs-induced Landau level splitting mechanism that is essentially the same as the one we discuss below. As emphasized above, we find that this picture is valid only if $C_3$ symmetry breaking is invoked. This is consistent with the more microscopic calculation presented in Ref.\ \onlinecite{Moon2012Energy}, which did not observe the experimental sequence for the parameter range suggested in Ref.\ \onlinecite{deGail2011Topologically} when all the emergent symmetries are kept. 
Finally, we also note that Ref.\ \onlinecite{Bi2019Designing}  suggested that layer-dependent strain could lead to the observed Landau level sequence near neutrality, although they did not perform an explicit calculation on this point.
The mechanism they considered, however, is different from the one discussed in the present work. In particular, in our model the $C_3$ symmetry does not differentiate the two layers. 
Very recently Ref.\ \onlinecite{Wu2019Identification} studied the effects of such strain on the pairing structure of the superconductor in TBG but that work did not address the Landau fans which are our interest here.

\section{Hofstadter Butterfly Calculations}
Our starting point is the single-valley continuum model for twisted bilayer graphene \cite{CastroNeto2007, bistritzer2011moire}. Let $q \equiv  \frac{4\pi}{3 a} \theta$ with $a = 2.46 $ \AA~being the lattice constant of monolayer graphene, and define the wave vectors $\mathbf q_1 = - q \hat{\mathbf y}$, $\mathbf q_2 = q (\sin \frac{2\pi}{3} \hat {\mathbf x} -\cos \frac{2\pi}{3}  \hat{\mathbf y})$, and  $\mathbf q_3 = q (-\sin \frac{2\pi}{3} \hat {\mathbf x} - \cos \frac{2\pi}{3}  \hat{\mathbf y})$. The Hamiltonian is given by
\begin{equation}\begin{split}\label{eq:HCont}
H = \left(
\begin{array}{cc}
\hbar v (-i \bm \nabla +  \mathbf q_{3}) \cdot \bm \sigma_{-\theta/2} & T(\mathbf r)\\
T^\dagger(\mathbf r) & \hbar  v (-i \bm \nabla - \mathbf q_{2}) \cdot \bm \sigma_{\theta/2} 
\end{array}
\right)
\end{split}\end{equation}
where $ \bm \sigma_{\phi} \equiv e^{i \phi \sigma_3 /2} \bm \sigma e^{-i \phi \sigma_3 /2}  $, $T(\mathbf r) = \sum_{j=1}^3  T^j  e^{- i \mathbf b_j \cdot \mathbf r  } $ with
\begin{equation}\begin{split}\label{eq:}
T^j = t_M \left(
\begin{array}{cc}
\alpha & e^{-i \frac{2\pi}{3} (j-1)}\\
e^{i  \frac{2\pi}{3} (j-1)} & \alpha
\end{array}
\right),
\end{split}\end{equation}
and $\mathbf b_j = \mathbf q_j - \mathbf q_1$ are reciprocal lattice vectors for the \moire potential. We use $v = 10^6$ m/s,  $t_M = 110$ meV \cite{bistritzer2011moire}, and $\alpha = 0.8$ to incorporate the effect of lattice relaxation \cite{koshino2018maximally}. 

A perpendicular magnetic field can be incorporated by substituting $-i \bm \nabla \mapsto -i \bm \nabla - e \mathbf A / \hbar$, with the vector potential $\mathbf A = -B y \, \hat{\mathbf x}$ in the Landau gauge. As in Ref.\ \onlinecite{Bistritzer2011PRB}, the spectrum for non-zero $B$ can be solved by first going into a Landau level basis for two decoupled layers, and then re-expressing the inter-layer potential $T(\mathbf r)$ in this basis (more details can be found in Appendix \ref{append:butterfly}). In practice, in the numerical calculations we keep only finite Landau levels (typically several hundreds) close to the charge neutrality. Note that in the limit $B\rightarrow 0$ an infinite number of Landau levels should to be kept, and so we only consider  fields $B > 1.2 $ T.

Eq.\ \eqref{eq:HCont} has all the effective symmetries of twisted bilayer graphene. In the following, we will also consider the effect of symmetry breaking (at the single-particle level). In particular, we will set $(1-\beta)T^1 = T^2 = T^3 $, such that $\beta \neq 0$ controls the degree of $C_3$ symmetry breaking. We will also introduce $C_2 $ symmetry breaking through $H_M = M_t \sigma_3 \oplus M_b \sigma_3$, which corresponds to a staggered chemical potential that would be present if the coupling to the  hBN substrate is significant.

We focus on small twist angles $\theta > 1.1^\circ$, slightly away from the true magic angle of $\theta\approx1.05^\circ$ \footnote{
Defined here as the angle at which the active band has a minimum band width, for the parameters we used.
}, for the following reasons: (1) In the experiment, the four-fold degeneracy is observed in a large range of angles, including in devices where the correlated insulators and superconductors are not observed. For instance, the sequence $\pm 4, \pm 8, \pm 12, \dots$ has been seen in the device at $1.27^\circ$ in Ref.\ \onlinecite{yankowitz2019tuning} under ambient pressure. This suggests that the key physics leading to the four-fold degeneracy should be in effect over a relative broad range of angles. (2) Due to the very small bandwidth exactly at the magic angle, the Landau level is not well-formed within our free-electron treatment, which complicates the analysis.

\section{Landau Fan for the Symmetric Case}
\begin{figure}[ht]
\centering
\includegraphics[width=0.48\textwidth]{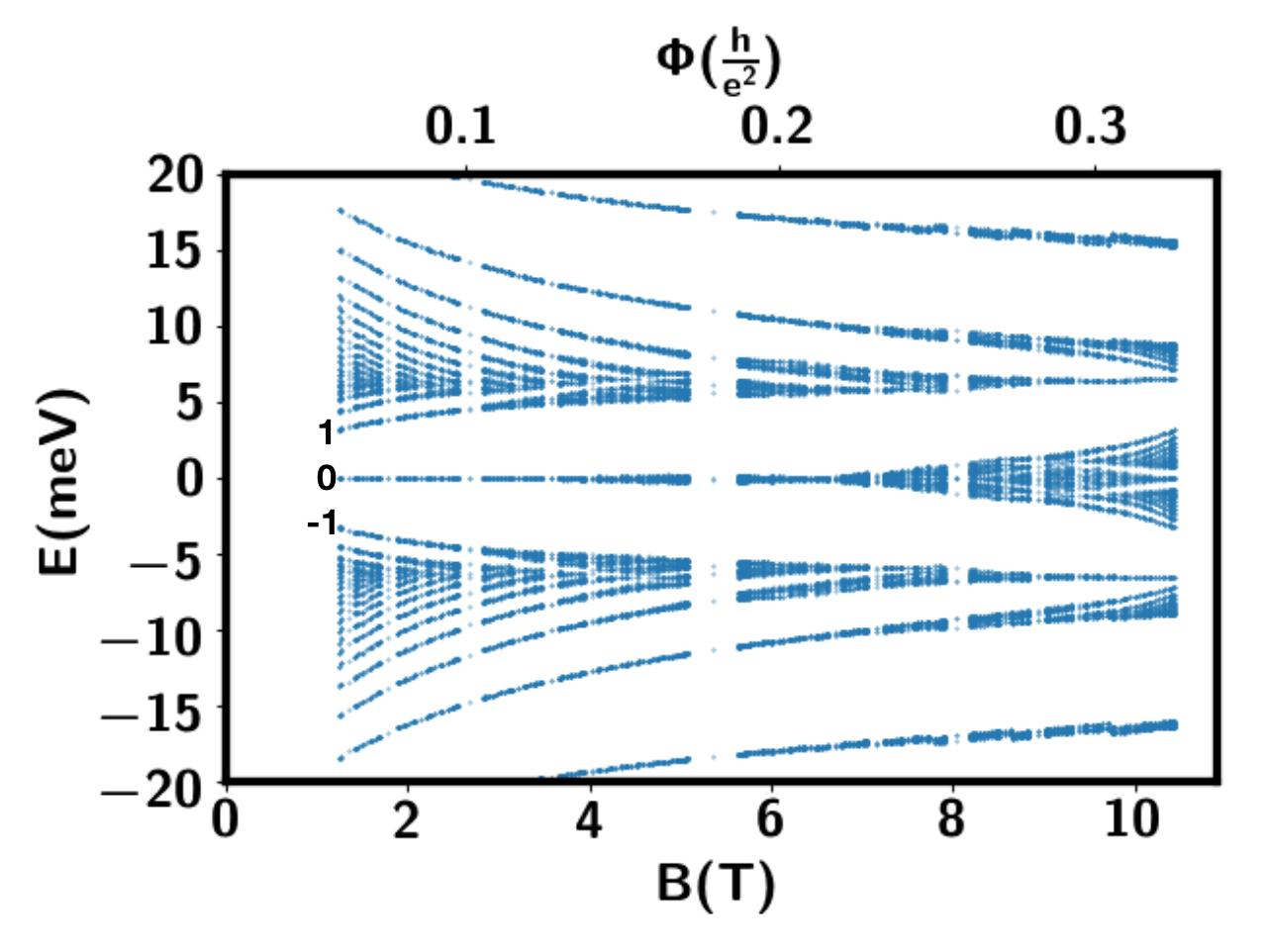}
\caption{Spectrum at twist angle $\theta=1.15 ^\circ$ with both $C_2$ and $C_3$ symmetry. From neutrality, only $n=0,\pm 1, \pm 2$ Landau levels can be resolved. Each of them is eight fold degenerate and the Landau fan sequence is $\pm 4, \pm 12, \pm 20$.
}
\label{fig:spectrum_strain_0}
\end{figure}

First, we show the results for the fully symmetric model, which retains the  $C_3$ and $C_2$ symmetries.  We will show that the Landau fan close to charge neutrality should have eight-fold degeneracy, in agreement with the experiments at twist angle $\theta=1.8^\circ$ but in disagreement with the results  near the magic angle \cite{cao2018correlated,yankowitz2019tuning}. The spectrum is shown in Fig.~\ref{fig:spectrum_strain_0}. From neutrality, only Landau levels $n=0,\pm 1, \pm 2$ are well-developed, and the higher levels are superseded by the Landau levels coming from the top and bottom of the active bands.
In a semi-classical picture, such blending of Landau levels occurs near the zero-field Van Hove singularity separating the equal-energy contours enclosing the two mini Dirac points from those enclosing the band extrema located at $\Gamma$.

Importantly, each Landau level emanating from charge neutrality is eight-fold degenerate. This degeneracy can be explained by noting the doubling due to spin, valley and layer degrees of freedom. We remark that the layer index is not generally a good quantum number; however, as these low-lying Landau levels can be understood as coming from the mini Dirac points, there is still a two-fold degeneracy arising from the mini-valley degeneracy. In contrast, in the experiments the Landau fan is only four-fold degenerate at such small twist angles.

\section{Landau Fan with $C_3$ Breaking}
In this section, we consider the effect of $C_3$ symmetry breaking by introducing an anisotropy $\beta$ such that $(1-\beta)T^1 = T^2 = T^3 $. We will not specify the physical origin of this anisotropy in this work. 
Our main purpose is to demonstrate that $C_3$ breaking can qualitatively change the Landau fan sequence, and so long as a mean-field description remains valid, we expect our conclusions to hold even if the $C_3$ breaking is interaction-driven.

\begin{figure}[ht]
\centering
\includegraphics[width=0.5\textwidth]{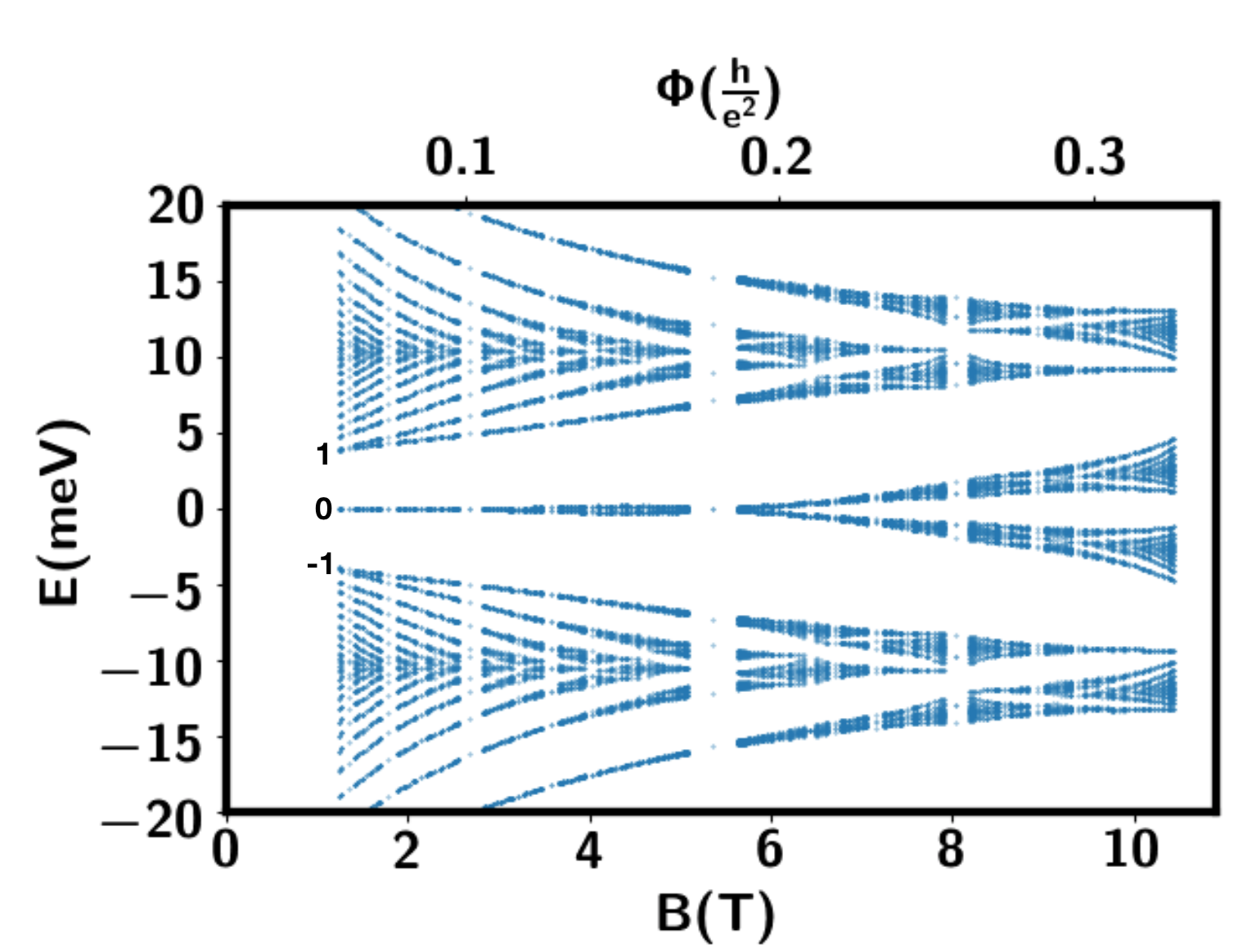}
\caption{Spectrum at twist angle $\theta=1.15 ^\circ$ with  $C_3$ breaking parameter $\beta = 0.07$.  Compared to the $C_3$ symmetric case, more Landau levels emerging from neutrality can be clearly resolved.
Meanwhile, in the magnetic field $B>1.2$ T, each Landau level is split into, each being four-fold degenerate. This leads to the Landan fan sequence $\pm 4, \pm 8, \pm 12,...$. $E_c$ is the energy of the first van-Hove singularity point discussed in Section IV.}
\label{fig:spectrum_strain_2}
\end{figure}

\begin{figure*}[ht]
\centering
  \begin{subfigure}[b]{0.48\textwidth}
    \includegraphics[width=\textwidth]{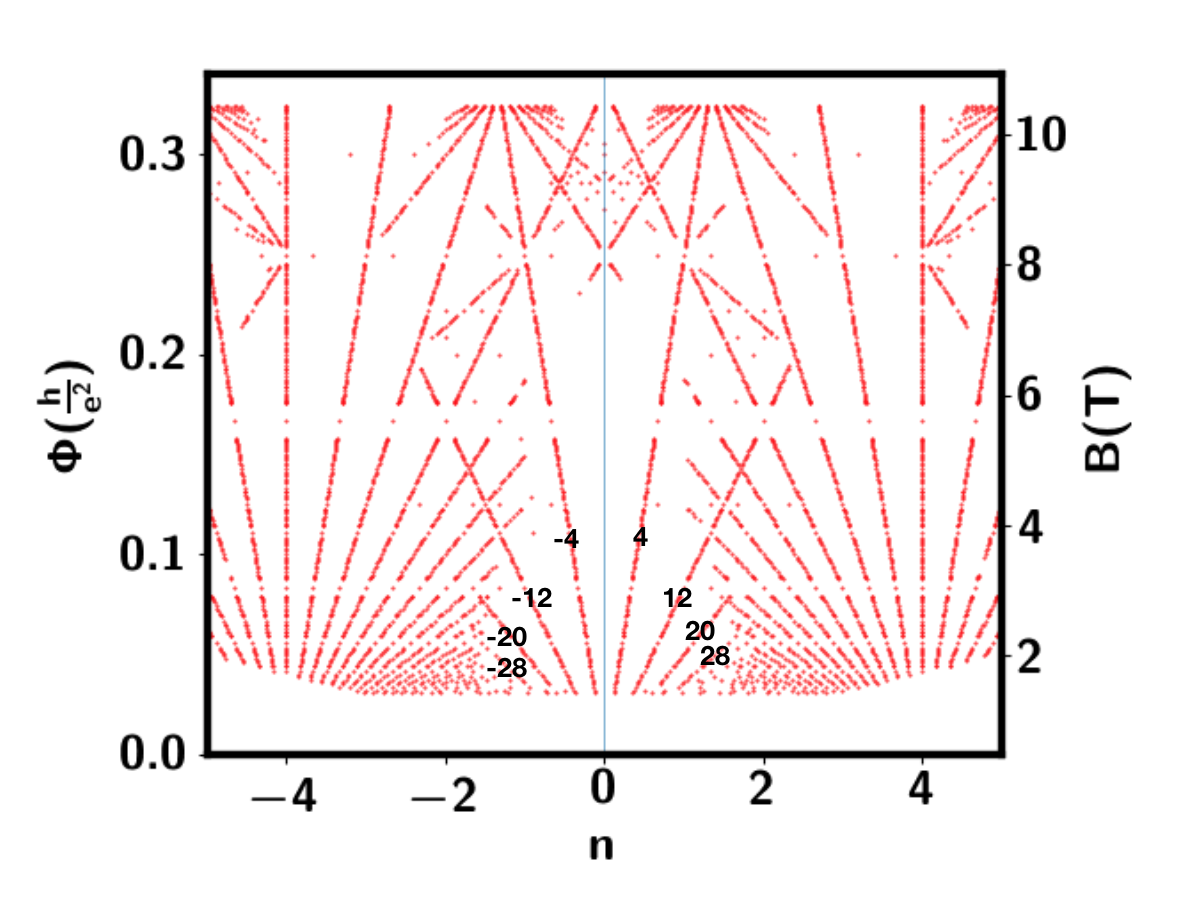}
    \caption{$C_3$ Symmetric.}
  \end{subfigure}
  \begin{subfigure}[b]{0.49\textwidth}
    \includegraphics[width=\textwidth]{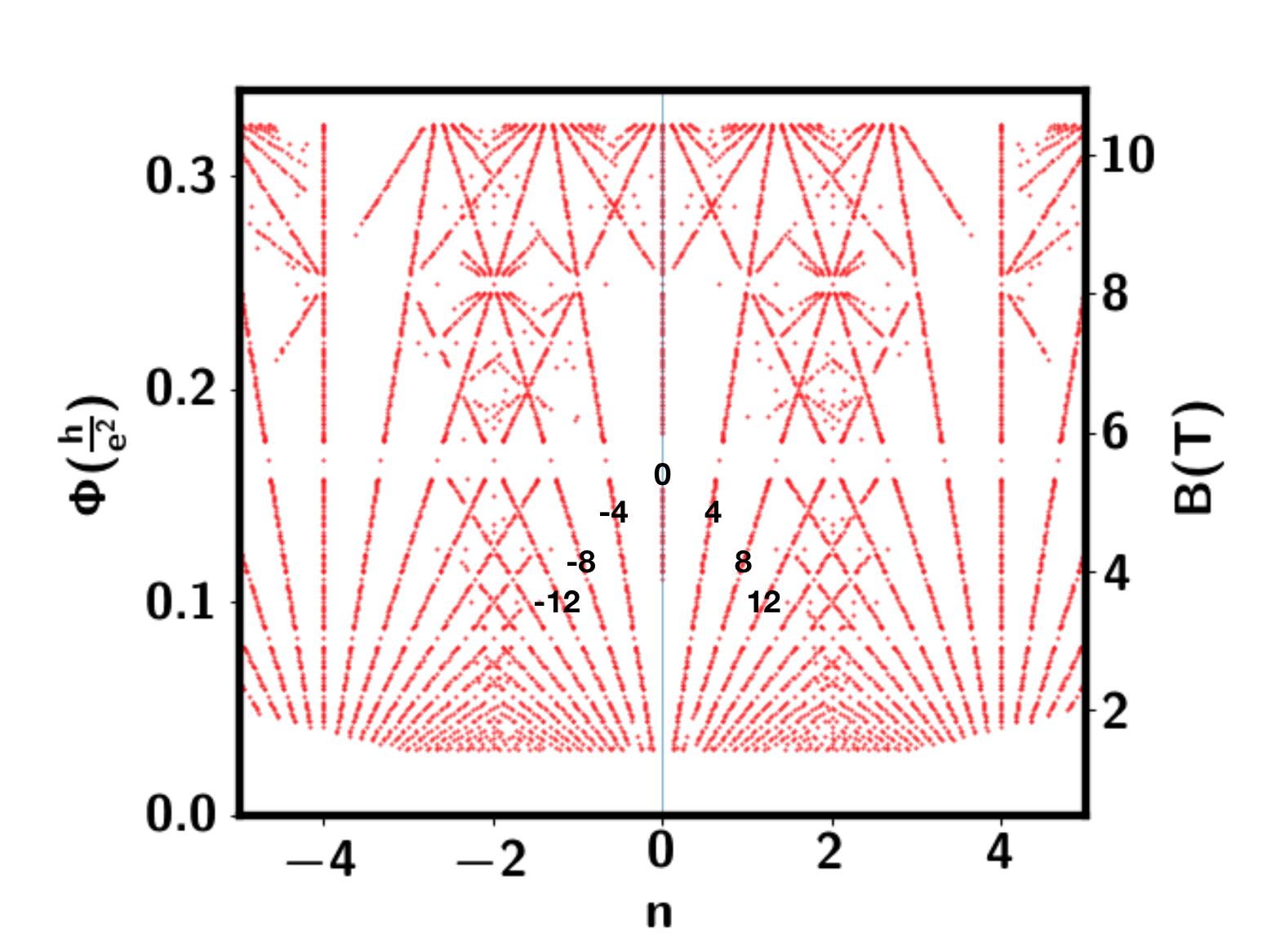}
    \caption{$C_3$ breaking parameter $\beta=0.07$}
  \end{subfigure}
  \caption{Wannier plot for twist angle $\theta=1.15^\circ$. With $C_3$ breaking, the experimentally observed sequence $\pm 4, \pm 8,\pm 12,...,$ is reproduced.}
  \label{fig:Wannier_plot}
\end{figure*}

The Landau fan spectrum  at $\beta=0.07$ is shown in Fig.~\ref{fig:spectrum_strain_2}. Compared to the $C_3$ invariant case, there are several clear differences. First, more Landau levels from neutrality can be seen, suggesting that the Van-Hove singularity is pushed further away from charge neutrality.
Such shift of Van-Hove singularity has  also been observed in STM experiment at neutrality\cite{choi2019imaging}. 
Second, the previously eight-fold degenerate Landau level is split into two groups when magnetic field is larger than a critical field $B_c(n)$ for each Landau level $n$.  $B_c(n=0)=6$ T and $B_c(n=\pm 1)=1.2$ T. For $|n|\geq 2$, $B_c$ is smaller than the lowest field we can reach in the calculation. Later we will argue that $B_c(n)\sim \frac{1}{|n|}$, which is confirmed for a smaller $C_3$ breaking parameter $\beta=0.01$ in the Appendix.
From the splitting of the Landau levels, one expects the Landau fan sequence $\pm 4, \pm 8, \pm 12, ...$ as observed in the experiments. This is  confirmed in the Wannier plot in Fig.~\ref{fig:Wannier_plot}.

\begin{figure}[h]
\begin{center}
{\includegraphics[width=0.48 \textwidth]{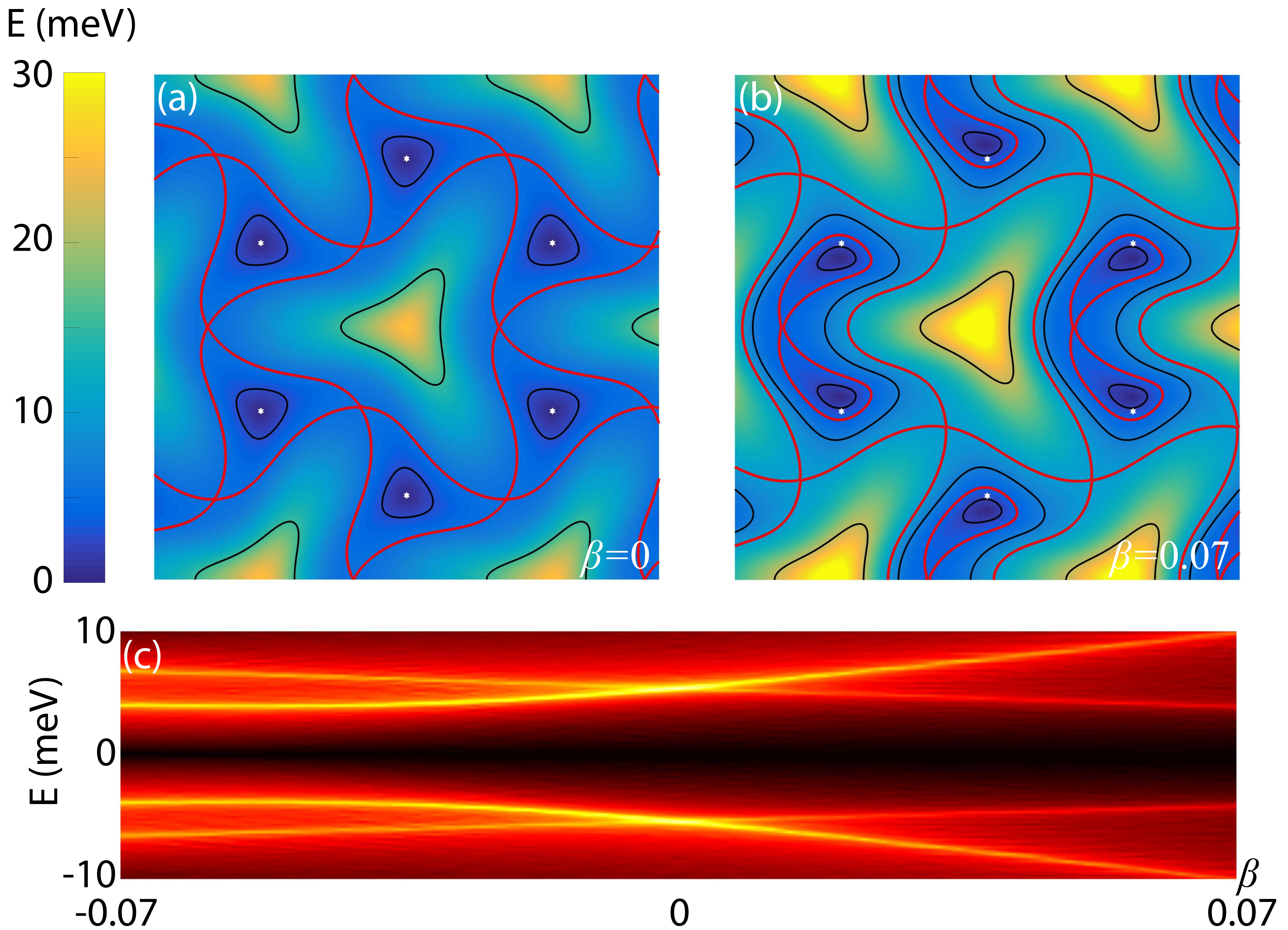}} 
\caption{Zero-field Van Hove singularities. (a) Representative equal energy contours (not equally spaced) of the conduction band for $\beta = 0$. The critical contour corresponding to the van Hove singularity is colored in red. The $K$ and $K'$ points are indicated by white asterisks. (b) Same as (a) but with $\beta = 0.07$. Note there are  two critical contours. (c) Density of states in arbitrary units. The bright yellow regions indicate the Van Hove singularities flanking charge neutrality, which splits as $\beta$ is tuned away from $0$.
\label{fig:zerofield}
 }
\end{center}
\end{figure}

Here we try to give a simple, qualitative explanation of the splitting of the eight-fold degenerate Landau level based on a semi-classical picture derived from the band structure at zero magnetic field.
With $C_2$ symmetry preserved, the four-fold degeneracy from spin and valley is robust. In our numerics, we have explicitly checked that the splitting happens within each fixed spin-valley sector, and so for the discussion here we can focus on a fixed sector.  
In the fully symmetric case (i.e., for $\beta = 0$), there are two Dirac cones at the $K$ and $K'$ points of the mini Brillouin zone (MBZ). At doping slightly above charge neutrality, the Fermi surface consists of independent electron pockets, each enclosing one of the Dirac points (Fig.\ \ref{fig:zerofield}a). As discussed in the previous section, the Landau levels arising from such orbits exhibit a two-fold mini-valley degeneracy, which, when combined with valley and spin degeneracy, leads to eight-fold degenerate Landau levels. 
When the doping is increased beyond the Van Hove singularity, the Fermi surface becomes a single hole pocket enclosing the $\Gamma$ point, which leads to four-fold degenerate Landau levels coming from the superlattice gap at the top of the active bands.

Importantly, the breaking of the $C_3$ symmetry unpins the Dirac cones from the $K$ and $K'$ points and distorts the dispersion. 
Similar to before, at doping slightly above charge neutrality, the Fermi surface consists of two independent electron pockets, and one expects mini-valley degeneracy in the corresponding Landau levels. However, at higher doping the mentioned distortion opens up the possibility of a new form of Fermi surface, which is electron-like and encloses {\it both} of the Dirac points (Fig.\ \ref{fig:zerofield}b). The corresponding Landau level would appear to emerge from charge neutrality, but would no longer showcase mini-valley degeneracy. In other words, the eight-fold degenerate Landau levels split. 
Equivalently, the emergence of such new form of Fermi surfaces can be seen from the splitting of the Van Hove singularity as $\beta$ increases (Fig.\ \ref{fig:zerofield}c). Such splitting from $C_3$ symmetry breaking is also discussed in Ref.\ \onlinecite{choi2019imaging}. Note that the preceding discussion assumes $\beta \geq 0$; while the Van Hove singularity still splits for negative values of $\beta$, the new equal energy contours are open orbits and do not lead to four-fold degenerate Landau levels in the semi-classical picture. Consistently, we find that the Landau levels do not split for $\beta <0$ (Appendix.\ref{appendix:negative_beta}).

As a complementary point of view, one can consider the Landau levels associated with the mini Dirac cones, which have energies $E_n =  v_{\rm eff}\sqrt{2 e \hbar n B}$ for $n\geq 0$. In this picture, the lifting of the mini-valley degeneracy originates from the hybridization between the two set of mini-Dirac Landau levels. Qualitatively, such hybridization is expected to become significant when $E_n \gtrapprox E_{c}$ \cite{deGail2011Topologically}, where $E_{c}$ denotes the energy of the first split Van Hove singularity arising from $C_3$ breaking. As such, we expect the Landau level splitting to be observable for $B \geq B_c(n)$ where $B_c(n) \propto 1/n$. This trend is in qualitative agreement with the numerics shown in Fig.\ \ref{fig:spectrum_strain_2} and Appendix \ref{append:largetwist}. However, we caution that such a simple picture does not explain, for instance, the splitting of the zeroth Landau level shown in Fig.\ \ref{fig:spectrum_strain_2}.

\section{Landau Fan with $C_2$ Breaking}

Recent experiments have seen signatures of $C_2$ breaking\cite{Aaron2019Emergent,lu2019superconductors}.  A typical $C_2$ breaking term is the staggered potential:
\begin{equation}
  H_{M}=M_t \psi_t^\dagger \sigma_z \psi_t+M_b \psi_b^\dagger \sigma_z \psi_b
\end{equation}
where $M_t,M_b$ is the staggered potential strength for top and bottom layer. $\sigma$ denotes Pauli matrices in the $A,B$ sublattice subspace.

If the sample is well-aligned with only the top h-BN substrate,  one  expects $M_t \neq 0, M_b \approx 0$\cite{zhang2019twisted,bultinck2019anomalous}.  In this case, the zeroth Landau level is always split into four groups, each of which has only two-fold spin degeneracy (see Append.~\ref{append:C_2}).  Therefore the first two Landau fan sequences are always $\pm 2, \pm 4$, which has indeed been observed in both Refs.~\onlinecite{Aaron2019Emergent} and \onlinecite{lu2019superconductors}. In this case, the activation gap for the $\nu=2$  quantum Hall sequence is proportional to the magnitude of $|M_t|-|M_b|$ and therefore is almost a constant when changing magnetic field. This feature may be a useful test for this scenario in the experiment. Note that the valley splitting for zeroth Landau level is from the valley-sublattice locking and should not be thought as a simple valley Zeeman coupling in the semi-classical picture.  To get a strong $\nu=2$ sequence in the single particle level, we find it necessary to invoke a strong mirror reflection breaking through a finite $|M_t|-|M_b|$. Therefore the sample in Ref.~\onlinecite{lu2019superconductors} should also have a strong mirror reflection symmetry breaking to explain the observed $\nu=\pm 2$ sequence. The origin of the mirror reflection symmetry breaking is not clear because no obvious signature of hBN alignment is reported in Ref.~\onlinecite{lu2019superconductors}.

\begin{figure*}[ht]
\centering
  \begin{subfigure}[b]{0.49\textwidth}
    \includegraphics[width=\textwidth]{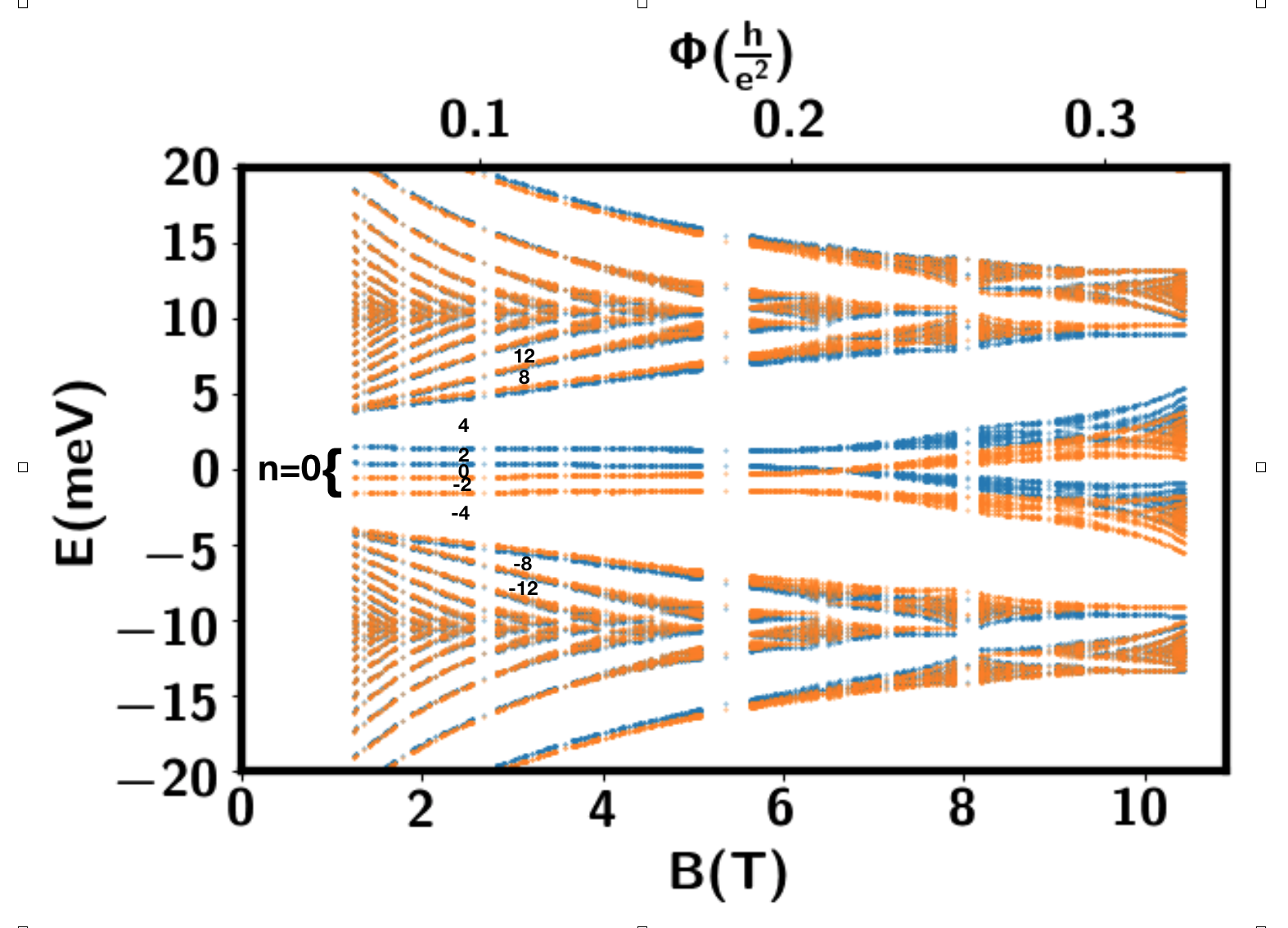}
  \end{subfigure}
  \begin{subfigure}[b]{0.49\textwidth}
    \includegraphics[width=\textwidth]{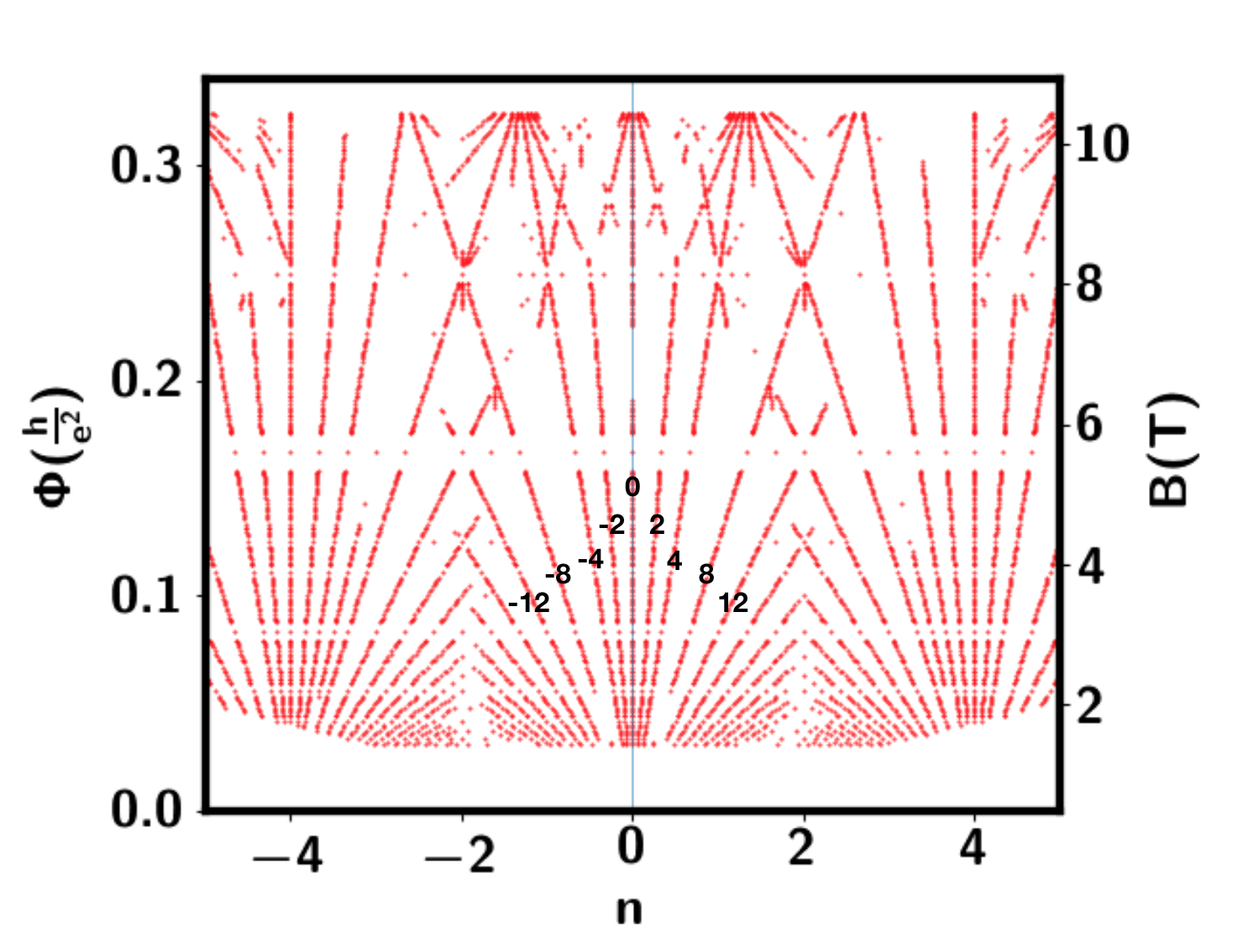}
  \end{subfigure}
  \caption{Spectrum and Wannier plot for $M_t=5$ meV, $M_b=0$, and $\beta=0.07$ at twist angle $\theta=1.15^\circ$. The two colors indicate states in the two microscopic valleys. With $C_3$ breaking, the experimentally observed sequence\cite{lu2019superconductors} $\pm 2, \pm 4, \pm 8,\pm 12,...,$ is reproduced.}
  \label{fig:plot_Mt_5_strain_1}
\end{figure*}

Next, we turn to the higher Landau levels. For $|n|\geq 1$ the valley splitting is quite small, as shown in Fig.~\ref{fig:plot_Mt_5_strain_1}.  Without $C_3$ breaking, the $|n| \geq 1$ Landau levels from mini $K$ and $K'$ overlap with each other and therefore  we still expect eight-fold degeneracy if $C_3$ is a good symmetry. $C_3$ breaking is necessary to get the observed four fold degeneracy in higher Landau levels. One example is illustrated in Fig.~\ref{fig:plot_Mt_5_strain_1}.

In the above, we just assume the order parameters without asking about their origins.  The simplest source to get $M_t=M$ and $M_b=0$ is just alignment with the top h-BN substrate. For spontaneous symmetry breaking, the most natural ansatz of spontaneous symmetry breaking in Hartree Fock calculation is actually $M_t=M_b$\cite{xie2018nature} and cannot reproduce the $\nu=\pm 2$ sequence.  It remains a question whether the $C_2$, mirror reflection and $C_3$ breakings needed to match the Landau fan in Ref.~\onlinecite{lu2019superconductors} are from external sources or from correlation effects.

\section{Discussion}
Since the first experiment  on near magic angle twisted bilayer graphene, the Landau fan near neutrality have been a mystery. The observed 4-fold Landau fan is in striking contrast with the naive theoretical expectation of an 8-fold degeneracy which is also what is seen at larger twist angles. The 4-fold degeneracy is seen in devices that approach but need not be too close to magic angle. In this paper we have shown how to understand these observations through explicit calculations of the Landau fan within the continuum model of the band structure. While there  exist previous such calculations with mixed success in explaining the experiments, we find that a key ingredient is to include $C_3$ symmetry breaking into the continuum Hamiltonian.  Upon approaching the magic angle the decreasing bandwidth (and the concomitant increase in density of states) leads to an amplified response of the electronic structure to a weak $C_3$ breaking anisotropy. This makes it easier for the equal energy contours near the mini-Dirac points to touch and reconstruct. Beyond a low energy scale the  equal energy contours then enclose both mini-Dirac points thereby reducing the Landau fan degeneracy from 8 to 4. We also consider the effect of $C_2$ breaking. 

With this understanding let us now revisit the experimental results on various devices. 
The samples in Ref.~\onlinecite{cao2018correlated,cao2018unconventional,yankowitz2019tuning}  behave as a semimetal at neutrality, suggesting a good $C_2 \mathcal T$ symmetry. To explain the four fold degeneracy of Landau fan, we require the existence of $C_3$ breaking, either from strain or from interaction driven symmetry breaking. Such $C_3$ breaking has been seen in STM experiments\cite{choi2019imaging,kerelsky2018magic} on other near magic angle twisted bilayer graphene samples. 

The sample in Ref.~\onlinecite{Aaron2019Emergent}  is aligned with the h-BN substrate on one side. This leads to an explicit breaking of both $C_2$  and the effective mirror (a $180^0$ rotation in 3d about an axis that is parallel to and bisects the 2 layers) symmetries. The $C_2$ breaking leads to to a gap at neutrality, consistent with the high resistivity observed at neutrality compared to the devices in Ref. \onlinecite{cao2018correlated,cao2018unconventional,yankowitz2019tuning}.  Extensive Landau fan data is not available in Ref. \onlinecite{Aaron2019Emergent} though the reported Landau levels  $\pm 2, \pm 4$ are reproduced by our analysis that takes into account both $C_2$ and mirror breaking. 

In the recent study of Ref.~\onlinecite{lu2019superconductors},  a number of additional features are reported. In contrast to Ref.~\onlinecite{cao2018correlated,cao2018unconventional,yankowitz2019tuning} but somewhat similar to Ref. \onlinecite{Aaron2019Emergent} charge neutrality is insulating with a sizeable gap $\approx 0.8$ meV. Further, a Landau fan sequence $\pm 2, \pm 4, \pm 8,......$ is reported near neutrality. The insulating behavior at neutrality could be due to $C_2 \mathcal T$ breaking either from explicit alignment with at least one of the h-BN substrates, or spontaneously due to interactions
\footnote{Alternately, it could be due to an interaction-driven inter-valley coherence  order\cite{po2018origin}.}.   In Ref. \onlinecite{lu2019superconductors} the observed Landau fan sequence near neutrality is $\pm 2, \pm 4, \pm 8, \dots$.   
In our calculations if we further retain the effective mirror symmetry, then we are unable to reproduce the observation of the $\pm 2$ level. Thus we conclude that the mirror symmetry must be strongly broken. 
An additional $C_3$ breaking is needed to further reduce the degeneracy of higher Landau levels to be four fold. 
Thus we think it likely that  $C_2, C_3$ and effective mirror are broken in the device in Ref. \onlinecite{lu2019superconductors}. 

 The $C_3$  and $C_2$ breaking  that we have crucially invoked in understanding the Landau fan clearly sets the stage on which the other correlated phenomena happen. An immediate open question is whether it is an explicit or spontaneous symmetry breaking driven by interactions. We hope to study this in the future.

\begin{acknowledgements}
We thank  Z.\ Bi, Y.\ Cao, P.\ Jarillo-Herrero, D.\ Mao, M.\ Yankowitz, and M.\ P.\ Zaletel for helpful discussions.
TS is supported by a US Department of Energy grant DE-SC0008739, and in
part by a Simons Investigator award from the Simons Foundation. 
HCP is supported by a Pappalardo Fellowship at MIT and a Croucher Foundation Fellowship.
\end{acknowledgements}

\bibliographystyle{apsrev4-1}
\bibliography{butterfly}

\onecolumngrid

\appendix

\hfill \break

\section{Calculation of Hofstadter butterfly \label{append:butterfly}}

In this section we describe the procedure of our calculation of Hofstadter butterfly spectrum. The method we used is essentially the same as in Ref.~\onlinecite{Bistritzer2011PRB}. 

We first describe a general scheme for any graphene moir\'e superlattice systems. Then calculation for the twisted bilayer graphene is straightforward.

For moir\'e system with large lattice constant, the two valleys always form separated spectrum and here we only treat valley $+$ as an example. At zero magnetic field, a moir\'e system is described by a continuum model:
\begin{equation}
  H=H_0+H_M
\end{equation}

$H_0$ is just the effective model for the valley:
\begin{equation}
  H_0=\sum_{\mathbf k} h_{\alpha \beta}(\mathbf k)f^\dagger_\alpha(\mathbf k)f_\beta(\mathbf k)
\end{equation}
where $\alpha,\beta$ is the combination of layer and sublattice index.

The $H_M$ is the moir\'e hopping or superlattice potential term. It involves the scattering of momentum from $\mathbf{k}$ to $\mathbf{k}+\mathbf{G_M}$, where $\mathbf{G_M}$ is the reciprocal vector of the small mini Brillouin zon (MBZ) folded by $H_M$.  A general $H_M$ term looks like
\begin{equation}
  H_M=\sum_{\mathbf k}\sum_{\mathbf{Q_j}}T_{\alpha \beta}(\mathbf{Q_j})f^\dagger_\alpha(\mathbf k)f_\beta(\mathbf{k+Q_j})
  \label{eq:moire_hamiltonian}
\end{equation}

When the moire\'e superlattice is formed by twisting two layers, the momenta of the Dirac cones (or more general band crossing points) from the two layers are different. Here we follow the notation of Ref.~\onlinecite{bistritzer2011moire} and define the momentum $\mathbf{k}$ in $\psi_{\alpha}(\mathbf{k})$ as the momentum relative to the Dirac point in the corresponding layer fixed by $\alpha$. In this convention $\mathbf{Q_j}$ does not belong to the reciprocal vector of the MBZ. 

The key idea to calculate the spectrum under magnetic field is the following. We first ignore $H_M$ and add magnetic field to $H_0$. We can easily get the new eigenstate in the Landau level basis. Then we express $H_M$ in the Landau level basis, which generically mix different Landau levels. Numerically we can still solve the resulting Hamiltonian and get the spectrum.

First let us solve $H_0$ plus magnetic field. The Hamiltonian is labeled by dynamical momenta $\pi_x=p_x-eA_x$ and $\pi_y=p_y-eA_y$ with the commutation relation:
\begin{equation}
  [\pi_x,\pi_y]=i \frac{1}{l_B^2} 
\end{equation}
where $l_B^2=\frac{\hbar}{eB}$.

We can define guiding center coordinates $R_x=x+l_B^2\pi_y$ and $R_y=y-l_B^2 \pi_x$ with commutation relation:

\begin{equation}
  [R_x,R_y]=-i  l_B^2 
\end{equation}

We also have $[\pi_a,R_b]=0$.  $\pi_a$ generates the Landau level index $n$ and $R_a$ generates the orbital within each Landau level. 

Define the ladder operator:
\begin{align}
a&=\frac{l_B}{\sqrt{2}}(\pi_x+i  \pi_y)\notag\\
a^\dagger&=\frac{l_B}{\sqrt{2}}(\pi_x-i \pi_y)
\end{align}

One can easily check that $[a,a^\dagger]=1$. Therefore $a$ can be viewed as an annihilation operator.  Landau level $n$ eigenstate satisfy:
\begin{equation}
  a\ket{n}=\sqrt{n} \ket{n}
\end{equation}

We use the Landau gauge: $\mathbf{A}=(-By,0)$, then $R_x=x+p_y l_B^2$ and $R_y=-p_x l_B^2$.  We can use $R_y=-p_x l_B^2$ to label the basis of each Landau level.  Each basis is then labeled by Landau level index $n$ and $k_x$.

We label the funciton $\varphi_{n,k_x}(y)$ as the wavefunction in the real space for the basis $\ket{n,k_x}$. The corresponding annihilation operator is labeled as $\psi_{\alpha;n,k_x}$, where $\alpha=A_1,B_1,A_2,B_2$ is the sublattice index.

The continuum model can be written in terms of the operator $f_{\alpha;k_x,ky}$, which is related to $\psi$ through
\begin{equation}
  f_{\alpha;k_x,k_y}=\sum_{n}  \tilde\varphi_{n,k_x}(k_y) \psi_{\alpha;n,k_x}
  \label{eq:landau_basis}
\end{equation}
where $\tilde\varphi_{n,k_x}(k_y)$ is the Fourier transformation of $\varphi_{n,k_x}(y)$ with respect to $y$.

We have the following translation property:
\begin{align}
\varphi_{n,k_x}(y)&=\varphi_{n,0}(y+k_x l_B^2)\notag\\
\tilde \varphi_{n,k_x}(k_y)&=\tilde \varphi_{n,0}(k_y)e^{i k_xk_y l_B^2}
\label{eq:translation_property}
\end{align}

$H_0$ can be expressed in  the following form
\begin{equation}
  H_0=\sum_{k_x}\sum_{nm;\alpha \beta}\psi^\dagger_{\alpha;n,k_x}h_{\alpha \beta; nm}\psi_{\beta;m,k_x}
\end{equation}
Here $k_x$ is defined in the whole  range of $R=[-\infty,\infty]$.  In the graphene problem, $h_{nm}$ is usually very simple. For example, in monolayer graphene, it is just an off diagonal term from $B,n$ to $A,n+1$.

$H_M$ generically scatters momentum. In terms of $\psi_{\alpha;n,k_x}$, $H_M$ in Eq.~\ref{eq:moire_hamiltonian} can be rewritten as
\begin{equation}
  H_M=\sum_{k_x}\sum_{\mathbf{Q}}\sum_{nm}T_{\alpha \beta}(\mathbf{Q})F_{nm;k_x}(\mathbf Q)\psi^\dagger_{\alpha;n;k_x}\psi_{\beta;m;k_x+Q_x}
  \label{eq:moire_hopping}
\end{equation}
where
\begin{align}
  F_{nm;k_x}(\mathbf Q)&=\sum_{k_y} \tilde \varphi^*_{n,k_x}(k_y)\tilde \varphi_{m,k_x+Q_x}(k_y+Q_y)\notag\\
  &=F^0_{nm}(\mathbf Q)e^{i(k_xQ_y+\frac{1}{2}Q_xQ_y)l_B^2}
  \label{eq:hopping_matrix_elements_derive}
\end{align}
where we have used Eq.~\ref{eq:translation_property}. We have
\begin{align}
  F^0_{nm}(\mathbf Q)&=\sum_{k_y}\tilde \varphi^*_{n,0}(k_y)\tilde \varphi_{m,0}(k_y+Q_y)e^{i k_y Q_xl_B^2}e^{\frac{i}{2} Q_xQ_y l_B^2}\notag\\
  &=\sqrt{\frac{m!}{n!}}(z_x+i z_y)^{n-m}e^{-\frac{|z|^2}{2}}L^{n-m}_m(|z|^2)
\end{align}
where $(z_x,z_y)=\frac{1}{\sqrt{2}}(Q_x,Q_y)$. $L^n_m(z)$ is the associated laguerre polynomials.

If we view $k_x=k^0_x+x \Delta$ with $x\in Z$ as site in a 1D chain,  Eq.~\ref{eq:moire_hopping} is basically a hopping term in this 1D chain.  The lattice constant is $\Delta=Q_x$ for the smallest non-zero $Q_x$. Here $\alpha,n$ is just the orbital label of each site labeled by $k_x$.  At each site $k_x$, the hopping term $t(k_x)=F_{nm;k_x}(Q_x)$ is a function of the site $k_x$. Generically we have a quasi-periodic 1D model because $t(x)\sim e^{ik_xQ_yl_B^2}\sim e^{i Q_x Q_y l_B^2 x}$ is oscillating according to Eq.~\ref{eq:hopping_matrix_elements_derive}. 

With the commensurate condition $\Delta Q_y l_B^2=2\pi \frac{q}{p}$, we have a unit cell with $p$ sites. Then we can solve the model with good quantum number $k_1=k_x^0\in [0,\Delta]$ and $k_2\in [-\frac{\pi}{p},\frac{\pi}{p}]$ as the crystal momentum of this 1D chain.  For TBG, the corresponding flux per moir\'e unit cell is $\Phi  = \frac{2\pi}{6} \frac{p}{q} \frac{h}{e^2}$.

Now we apply the above general scheme to the TBG. We keep $M$ Landau levels. Typical $M$ is several hundreds.  For the flux corresponding to $\Phi  = \frac{2\pi}{6} \frac{p}{q} \frac{h}{e^2}$,  we can construct a $4pM$ matrix for $H_0+H_M$ corresponding to each fixed $(k_1,k_2)$. We use $f_{\alpha,n,j}(k_1,k_2)$ to label the state with momentum $k_2$ in the 1D chain fixed by $k_1$. $j=1,2,...,p$ is the sublattice index for each unit cell.

The Hamiltonian corresponding to $H_0$ is:
\begin{align}
  H_0&=-\frac{\sqrt{2}\upsilon}{l_B} \sum_{\mathbf k}\sum_{n=0,1,...M}\big( e^{i\frac{\theta}{2}}\sqrt{n+1}f^\dagger_{A_1;n+1;j}(\mathbf k)f_{B_1;n;j}(\mathbf k)+h.c.\notag\\
  &+e^{-i\frac{\theta}{2}}\sqrt{n+1}f^\dagger_{A_2;n+1;j}(\mathbf k)f_{B_2;n;j}(\mathbf k)+h.c. \big)\notag\\
  &+\sum_{\mathbf k}\sum_n M_t (f^\dagger_{A_1;n,j}(\mathbf k)f_{A_1;n,j}(\mathbf k)-f^\dagger_{B_1;n,j}(\mathbf k)f_{B_1;n,j}(\mathbf k)) \notag\\
  &+\sum_{\mathbf k}\sum_n M_b (f^\dagger_{A_2;n,j}(\mathbf k)f_{A_2;n,j}(\mathbf k)-f^\dagger_{B_2;n,j}(\mathbf k)f_{B_2;n,j}(\mathbf k)) 
  \label{eq:final_H0}
\end{align}
Note that here $\mathbf{k}=(k_1,k_2)$ is just a label of the eigenstate of the 1D chain. It is not the true 2D momentum.

For TBG, $H_M$ is
\begin{align}
  H_M&=\sum_{\mathbf k}\sum_{nm}\big(F^0_{nm}(\mathbf Q^0)e^{ik_x^0Q^j_yl_B^2}e^{i \Delta Q^j_y l_B^2j}
  +\sum_{j=1,2} e^{i k_2}F^0_{nm}(\mathbf Q^j)e^{i(k_x^0Q^j_y+\frac{1}{2}Q^j_xQ^j_y)l_B^2}e^{i \Delta Q^j_y l_B^2j} \big)
  f^\dagger_{\alpha; n;j}(\mathbf k) T^j_{\alpha \beta}f_{\beta;m;j}(\mathbf k)
\end{align}
where $T^j_{\alpha \beta}$ is the inter-layer moir\'e hopping matrix.  We have $T^j_{tb}=T^{j\dagger}_{bt}=T^j$. Here $t,b$ means top and bottom layer. The momentum transfer is $\mathbf{Q_1}=(0,\frac{4\pi}{3 a_M})$, $\mathbf{Q_2}=(-\frac{2\pi}{\sqrt{3}a_M},-\frac{2\pi}{3 a_M})$ and $\mathbf{Q_3}=(\frac{2\pi}{\sqrt{3}a_M},-\frac{2\pi}{3 a_M})$.

\begin{equation}
  T^j=t_M\left(\begin{array}{cc}
  \alpha & e^{-i \varphi (j-1)}\\
  e^{i\varphi (j-1)} &\alpha
  \end{array}
  \right)
\end{equation}
where $\varphi=\frac{2\pi}{3}$. $\alpha\leq 1$ is a parameter to incorporate lattice relaxation. We use $\alpha=0.8$ in our calculation.

\section{Landau levels when $\beta<0$ \label{appendix:negative_beta}}
\begin{figure}[h]
\centering
  \begin{subfigure}[b]{0.45\textwidth}
    \includegraphics[width=\textwidth]{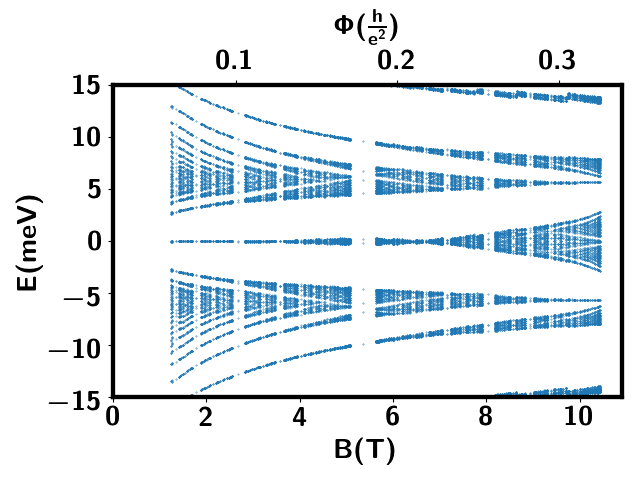}
  \end{subfigure}
  \begin{subfigure}[b]{0.45\textwidth}
    \includegraphics[width=\textwidth]{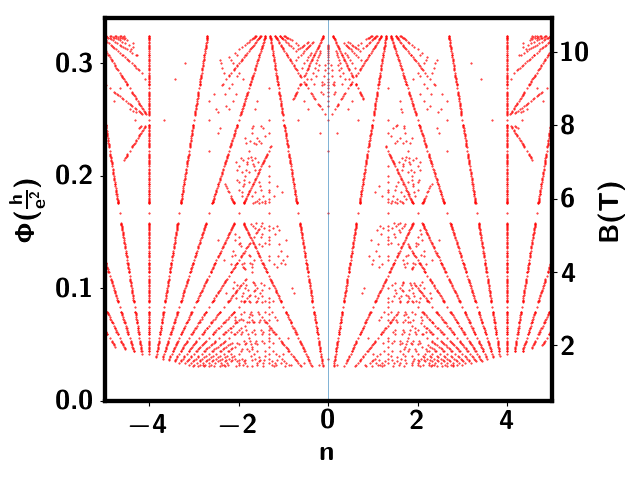}
  \end{subfigure}
  \caption{The Landau fan with $C_3$ breaking parameter $\beta=-0.03$ at twist angle $\theta=1.15^\circ$. $C_2$ symmetry is preserved. In contrast to positive $\beta$, there is no splitting of the Landau levels and the Landau fan is eight fold degenerate.}
  \label{fig:negative_beta}
\end{figure}

In the main text we show that a $C_3$ breaking with $\beta>0$ can reduce the degeneracy to four fold. Here we discuss the case with $\beta<0$. As is shown in Fig.~\ref{fig:negative_beta}, each Landau level from the charge neutrality is still eight fold degenerate for $\beta<0$. This implies that very specific form of $C_3$ is required to explain the experimental result of four fold degenerate Landau fan.

\section{Landau fan at larger twist angle
\label{append:largetwist}}

\begin{figure}[h]
\centering
  \begin{subfigure}[b]{0.32\textwidth}
    \includegraphics[width=\textwidth]{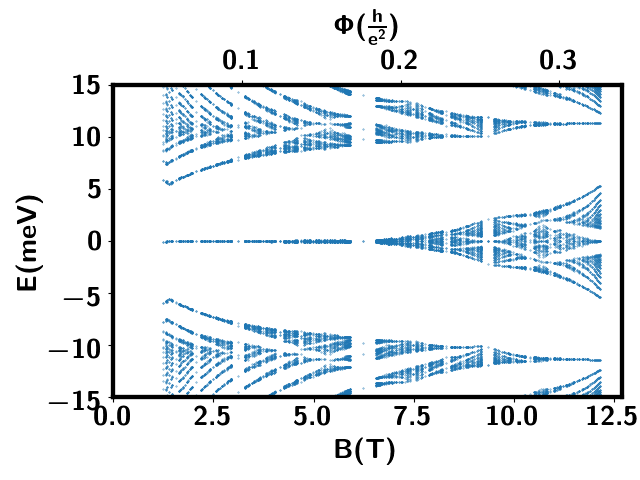}
    \caption{$C_3$  symmetric}
  \end{subfigure}
  \begin{subfigure}[b]{0.32\textwidth}
    \includegraphics[width=\textwidth]{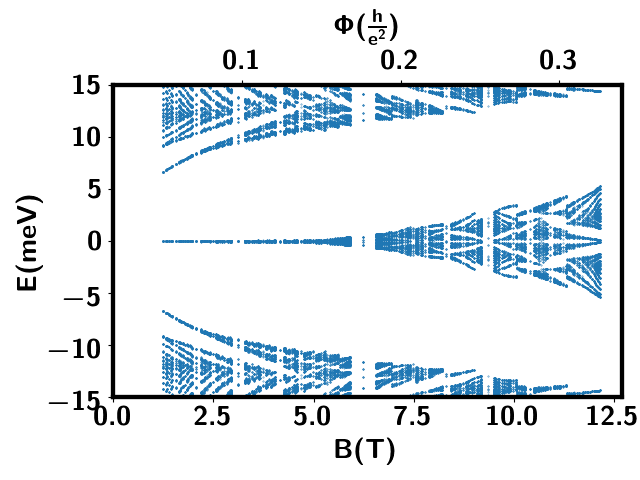}
    \caption{$C_3$ breaking parameter $\beta=0.03$}
  \end{subfigure}
  \begin{subfigure}[b]{0.32\textwidth}
    \includegraphics[width=\textwidth]{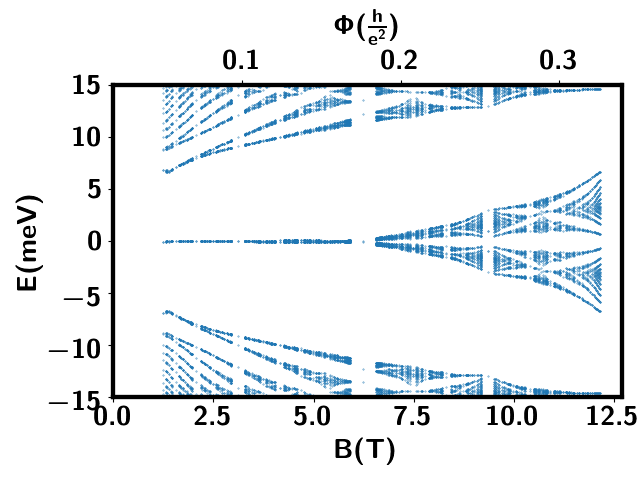}
    \caption{$C_3$ breaking parameter $\beta=0.07$}
  \end{subfigure}
  \caption{Hofstadter butterfly spectrum at $\theta=1.24^\circ$ for various $C_3$ breaking parameters. $C_2$ symmetry is preserved. For $\beta=0.07$, one can see obvious splitting of $|n| \geq 1$ Landau levels when energy is larger than $8$ meV. Correspondingly, the splitting happens above a critical magnetic field $B_c(n=1)\approx 1.2$ T and $B_c(n=2)\approx 2.4$ T. They satisfy the scaling $B_c(n)\sim \frac{1}{|n|}$ derived from semiclassical picture in the main text. }
  \label{fig:spectrum_larger_angle}
\end{figure}

\begin{figure}[h]
\centering
  \begin{subfigure}[b]{0.32\textwidth}
    \includegraphics[width=\textwidth]{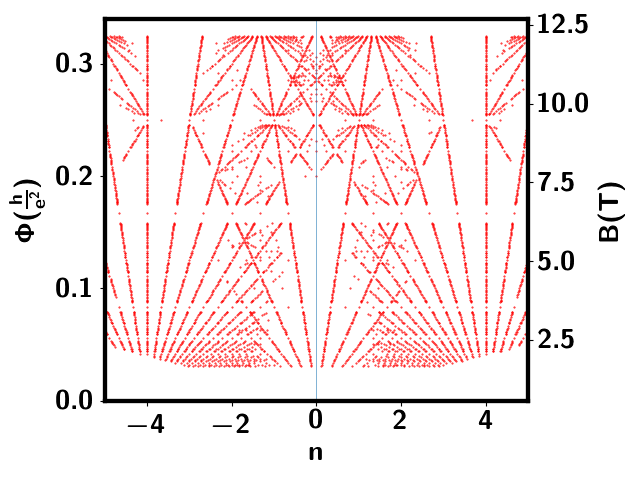}
    \caption{$C_3$ symmetric}
  \end{subfigure}
  \begin{subfigure}[b]{0.32\textwidth}
    \includegraphics[width=\textwidth]{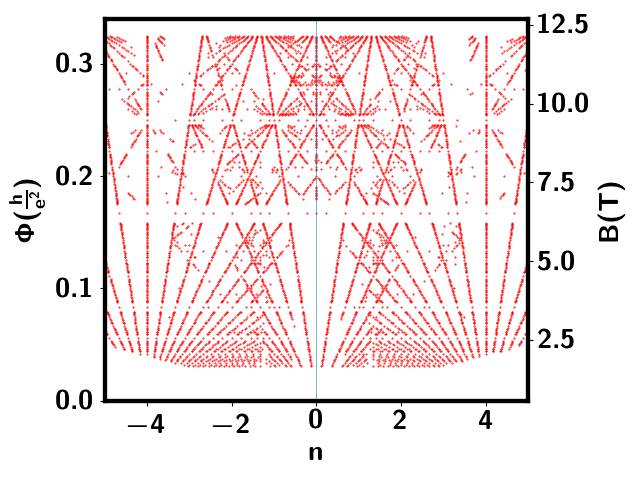}
    \caption{$C_3$ breaking parameter $\beta=0.03$}
  \end{subfigure}
  \begin{subfigure}[b]{0.32\textwidth}
    \includegraphics[width=\textwidth]{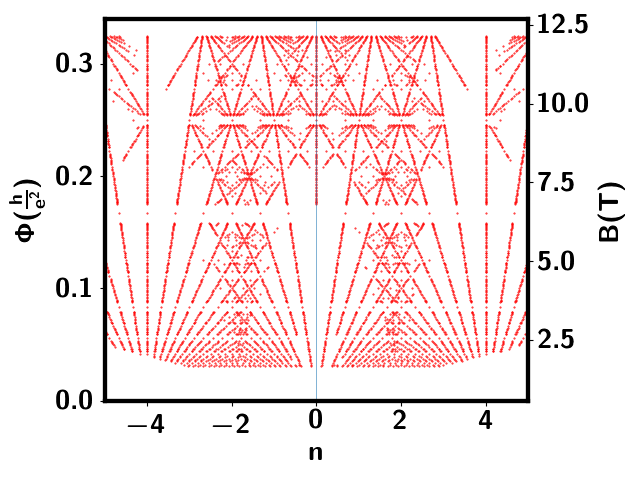}
    \caption{$C_3$ breaking parameter $\beta=0.07$}
  \end{subfigure}
  \caption{Wannier plot at $\theta=1.24^\circ$ for various $C_3$ breaking strength. $C_2$ symmetry is preserved. With $\beta=0.07$, we can still see the sequence $n=\pm 8 \Phi$ at magnetic field $B>2.5$ T. }
  \label{fig:wannier_plot_larger_angle}
\end{figure}

For $\theta=1.24^\circ$, the spectrum and the Wannier plots are shown in  Fig.~\ref{fig:spectrum_larger_angle} and Fig.~\ref{fig:wannier_plot_larger_angle}. Clearly even at this larger angle $C_3$ breaking can lead to four fold degenerate Landau fan at finite magnetic field, even though the band width is as large as $40$ meV here.

\section{Landau Fan with $C_2$ Breaking\label{append:C_2}}

Recent experiments have seen signatures of $C_2$ breaking\cite{Aaron2019Emergent,lu2019superconductors}.  A typical $C_2$ breaking term is just the staggered potential:
\begin{equation}
  H_{C_2}=M_t \psi_t^\dagger \sigma_z \psi_t+M_b \psi_b^\dagger \sigma_z \psi_b
\end{equation}
where $M_t,M_b$ is the staggered potential strength for top and bottom layer. $\sigma$ is a Pauli matrix in the $A,B$ sublattice subspace.

For $C_2$ breaking coming from alignment with the top (bottom) h-BN layer, we have $M_t=M,M_b=0$ ($M_t=0,M_b=M$). 

For $C_2$ breaking from spontaneous symmetry breaking, a natural choice is $M_t=M_b=M$\cite{xie2018nature}.  Alignment with both top and bottom h-BN layers also gives this term.

In the following we discuss the Landau fan close to neutrality from the above two different $C_2$ breaking ansatz.

\subsection{Zeroth Landau levels \label{append:zero_landau_level_splitting}}
We start from the $n=0$ Landau level with an emphasis on the observed $\nu=\pm 2$ sequence\cite{Aaron2019Emergent,lu2019superconductors}. When $C_2$ is preserved, there are two Dirac cones at $K$ and $K'$ of the MBZ for each spin-valley sector. 

Let us suppress the spin index. For the Dirac cone at $K$ (or $K'$) of valley $a=\pm$, we can write a $2\times 2$ effective model:
\begin{equation}
  H_{eff}=\sum_{\mathbf k} \psi^\dagger(\mathbf k)\left(\begin{array}{cc} M_{K,K'} & \upsilon_{eff} (k_x \mp i k_y) \\ \upsilon_{eff}(k_x \pm i ky) & -M_{K,K'}  \end{array}\right)\psi(\mathbf k)
\end{equation}
where $\psi(\mathbf k)$ is a two component spinor. Note that the pseudospin basis of $\psi(\mathbf k)$ is a mixture of the sublattice and layer.

\begin{figure}[H]
\centering
\begin{subfigure}[b]{0.49\textwidth}
    \includegraphics[width=\textwidth]{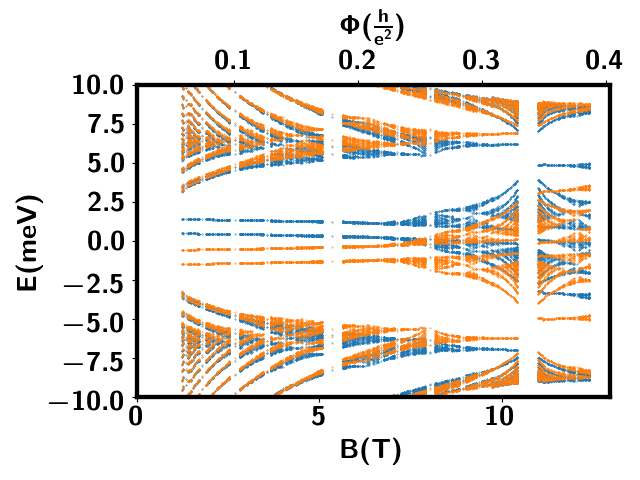}
    \caption{$M_t=5$ meV, $M_b=0$}
  \end{subfigure}
  \begin{subfigure}[b]{0.49\textwidth}
    \includegraphics[width=\textwidth]{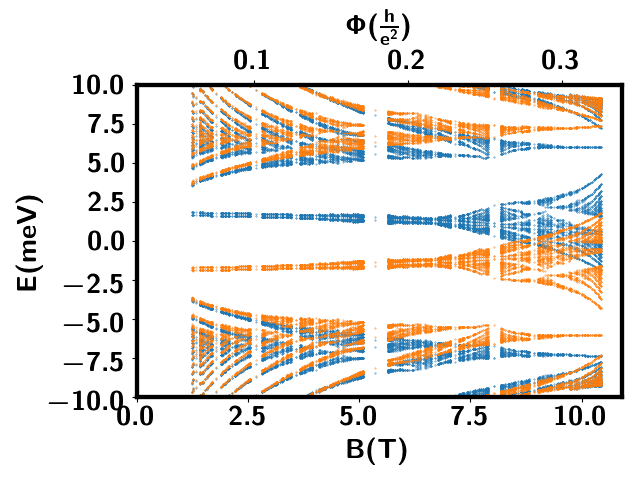}
    \caption{$M_t=5$ meV, $M_b=4$ meV}
  \end{subfigure}
  \caption{Spectra at twist angle $1.15^\circ$ for two different $C_2$ breaking ansatz.  $C_3$ symmetry is preserved. The two colors indicate states in the two microscopic valleys. The zeroth Landau levels are split to four or two groups and do not disperse with magnetic field up to $8$ Tesla.  For the second plot with a small $|M_t|-|M_b|$,  the splitting between $K$ and $K'$ state within each valley is very small. In this case, the $\nu=\pm 2$ quantum Hall sequence should be absent without invoking interaction effects.}
  \label{fig:spectrum_C2}
\end{figure}

When $M_t=M_b$, mirror reflection guarantees $M_K=M_{K'}$.  When $|M_t|>>|M_b|$, we also expect $|M_K|>>|M_{K'}|$.  Let us make $M_t=M_b=0$ first. Now we add magnetic field and get four zeroth Landau levels formed by $(\pm ,K/K')$ for each spin sector.  In total the zeroth Landau level has eight fold degeneracy.  Next we treat $M_t$ and $M_b$ as small perturbation and add them on this eight fold degenerate $n=0$ Landau level subspace. For the zeroth Landau level, it is well known that the pseudospin is polarized and locked to the valley.  The $+,K$ and $+,K'$  eighenstate only contains $\psi_1$ component while the $-K$ and $-K'$  eigenstate only contains the component $\psi_2$. Because of this special property, $M_K$ and $M_{K'}$ (resulting from $M_t$ and $M_b$) just shift the energy of the zeroth Landau level from $(\pm ,K)$ ($(\pm, K')$) by a constant  energy $\pm M_{K}$ ($\pm M_{K'}$).  The final result is that we have four groups of $n=0$ Landau levels: two groups from valley $-$ at  energy $-M_{K}$ and $-M_{K'}$, two groups from valley $+$ at energy $M_{K'}$ and $M_{K}$. This is clearly shown in Fig.~\ref{fig:spectrum_C2}. For $M_t=5$ meV, $M_b=0$, each of the four groups only has two fold spin degeneracy. There should be a $\nu=\pm 2$ Quantum Hall sequence with the activation gap almost constant with the magnetic field.  For $|M_t|\approx |M_b|$, we have $|M_K|\approx |M_{K'}|$. Therefore there are only two groups of $n=0$ Landau levels.  In this case the $\nu=\pm 2$ quantum Hall sequence should be quite weak or even absent without invoking interaction effects.

\subsection{Higher Landau Levels}

\begin{figure}[H]
\centering
\begin{subfigure}[b]{0.32\textwidth}
    \includegraphics[width=\textwidth]{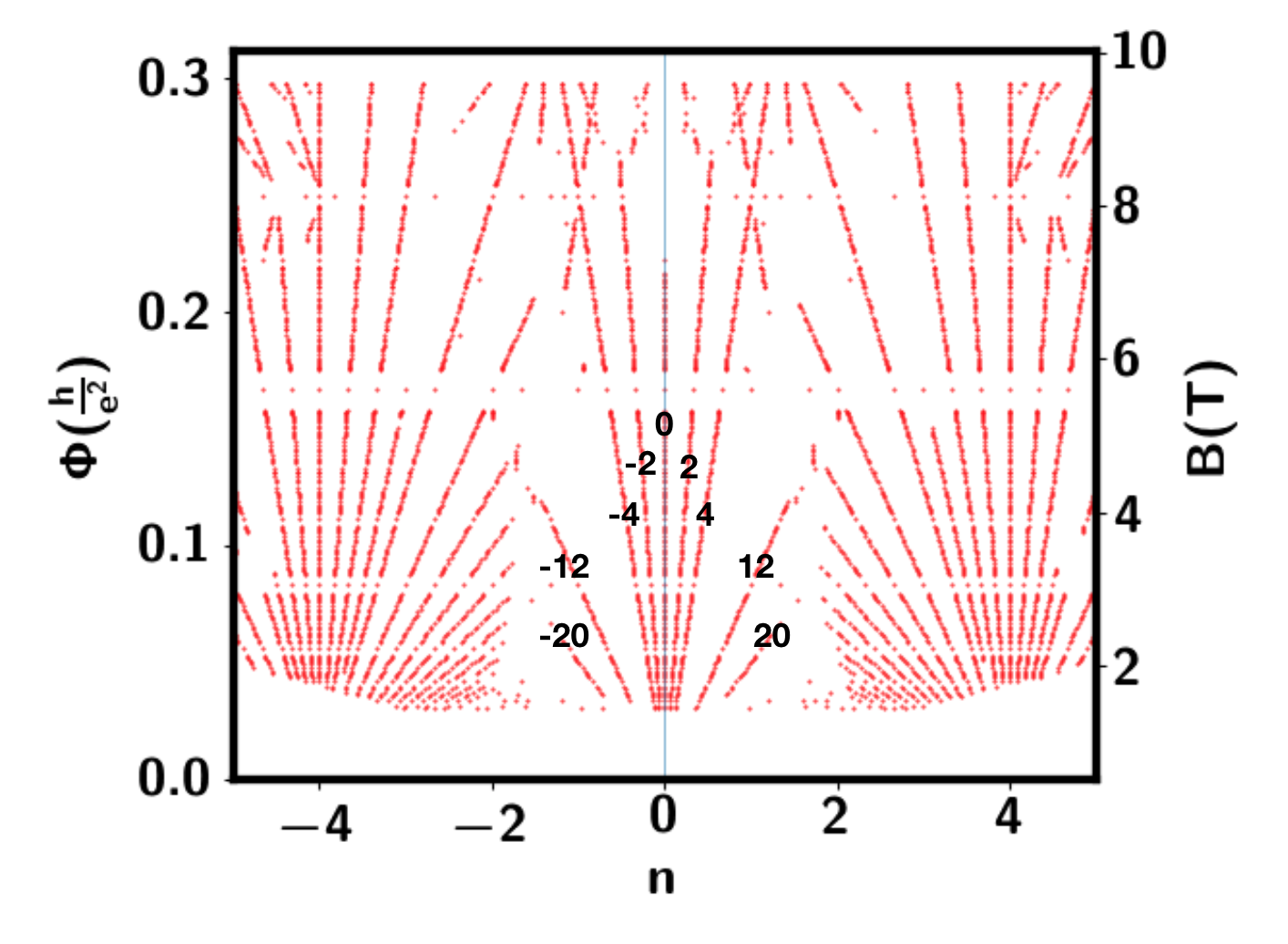}
    \caption{$M_t=5$ meV, $M_b=0$; $C_3$ symmetric}
  \end{subfigure}
  \begin{subfigure}[b]{0.32\textwidth}
    \includegraphics[width=\textwidth]{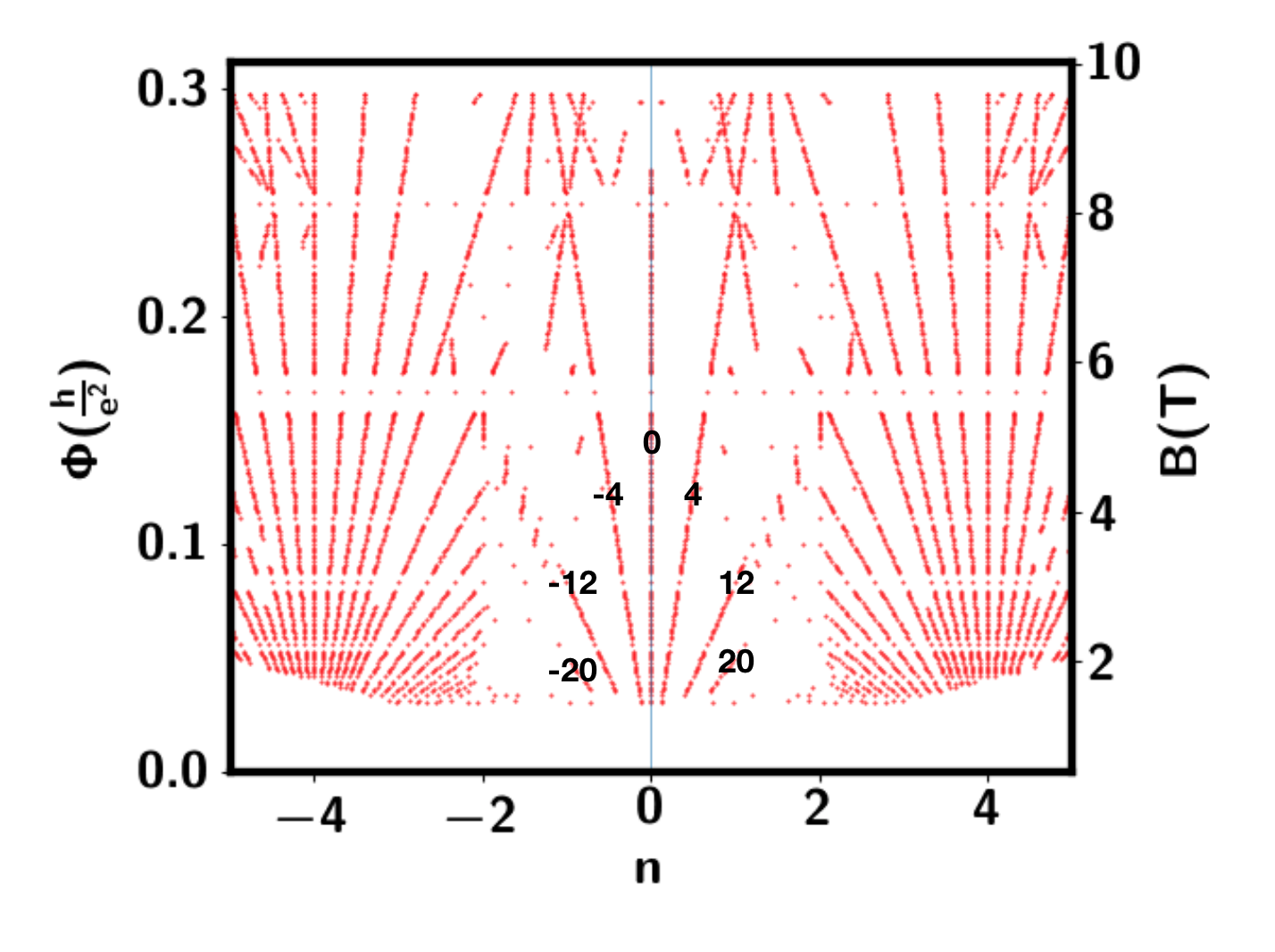}
    \caption{$M_t=M_b=5$ meV; $C_3$ symmetric}
  \end{subfigure}
    \begin{subfigure}[b]{0.32\textwidth}
    \includegraphics[width=\textwidth]{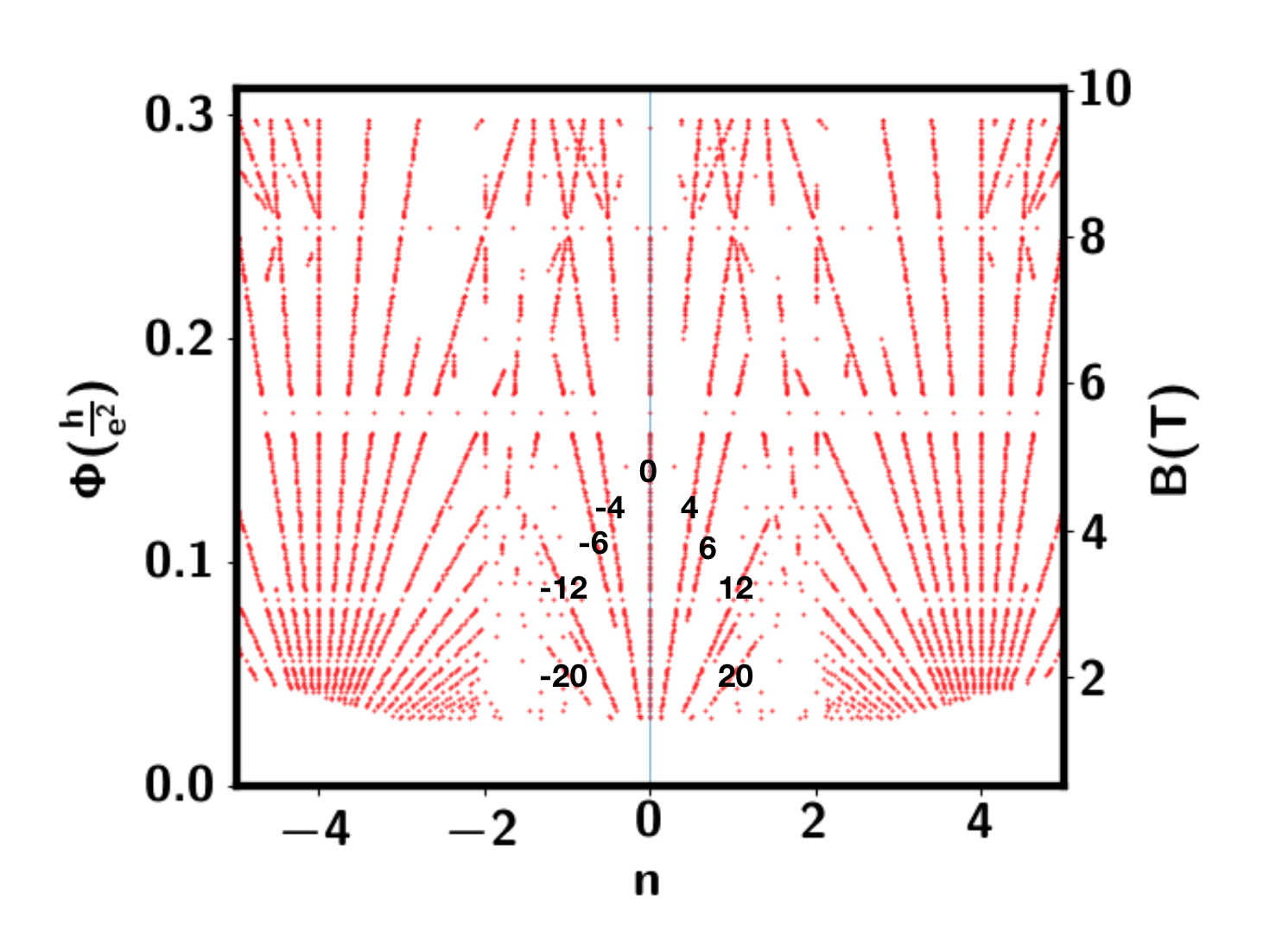}
    \caption{$M_t=M_b=5$ meV; $C_3$ breaking parameter $\beta=0.03$.}
  \end{subfigure}
  \caption{Wannier plots for several $C_2$ and $C_3$ breaking ansatz at twist angle $\theta=1.15^\circ$. None of these fully reproduce the Landau fan sequence $\pm 2, \pm 4, \pm 8, \pm 12 ,...$ in Ref.~\onlinecite{lu2019superconductors}. }
  \label{fig:wannier_C2}
\end{figure}

For higher Landau levels, Ref.~\onlinecite{lu2019superconductors} observe four fold degeneracy. Here we try to give an explanation with $C_2$ breaking ansatz.  With $C_2$ breaking, the only exact degeneracy at finite magnetic field is the two fold spin degeneracy (we ignore the small spin Zeeman coupling). However, for $n\geq 1$, the valley splitting is quite weak, as shown in Fig.~\ref{fig:spectrum_C2}. For $M_t=5$ meV and $M_b=0$, the mirror reflection is strongly broken. However, the  $n\geq 1$ Landau levels from $K$ and $K'$  for each valley still overlap with each other. Therefore the $n\geq 1$ Landau levels have eight fold degeneracy as long as $C_3$ is not broken. With $C_3$ breaking, the Landau levels from $K$ and $K'$ will couple with each other and reduce the degeneracy to four fold, as discussed for $C_2$ symmetric case. We show the Wannier plots for several cases in Fig.~\ref{fig:wannier_C2}.  One can see that the ansatz with $M_t=M_b$ can not give the $\nu=\pm 2$ quantum Hall sequence regardless of whether $C_3$ is broken. The only ansatz  which  can reproduce the sequence $\pm 2, \pm 4, \pm 8, \pm 12,...$ in Ref.~\onlinecite{lu2019superconductors} is the one shown in Fig.~\ref{fig:plot_Mt_5_strain_1} in the main text.

\end{document}